# A Natural Topological Insulator

P. Gehring[1]*, H. M. Benia[1], Y. Weng[2], R. Dinnebier[1], C. R. Ast[1], M. Burghard[1] and K. Kern[1,3]

[1]Max-Planck-Institut für Festkörperforschung, Heisenbergstrasse 1, D-70569 Stuttgart (Germany)

[2]Institut für Materialwissenschaft, Universität Stuttgart, Heisenbergstrasse 3, D-70569 Stuttgart (Germany)

[3]Institut de Physique de la Matière Condensée, Ecole Polytechnique Fédérale de Lausanne, CH-1015 Lausanne (Switzerland)

ABSTRACT

The earth crust and outer space are rich sources of technologically relevant materials which have found application in a wide range of fields. Well-established examples are diamond, one of the hardest known materials, or graphite as a suitable precursor of graphene. The ongoing drive to discover novel materials useful for (opto)electronic applications has recently drawn strong attention to topological insulators. Here, we report that Kawazulite, a mineral with the approximate composition $Bi_2(Te,Se)_2(Se,S)$, represents a naturally occurring topological insulator whose electronic properties compete well with those of its synthetic counterparts. Kawazulite flakes with a thickness of a few tens of nanometers were prepared by mechanical exfoliation. They exhibit a low intrinsic bulk doping level and correspondingly a sizable mobility of surface state carriers of more than 1000 cm$^2$/Vs at low temperature. Based on these findings, further minerals which due to their minimized defect densities display even better electronic characteristics may be identified in the future.

The extraordinary properties and architecture of naturally occurring compounds have inspired the synthesis of numerous biomimetic materials, a field that has attracted tremendous interest in the past few years.[1] Nature's ability to serve as a valuable blueprint for modern synthetic compounds is owed to the fact that it houses a wide variety of materials which have been forged over geological periods of time under extreme conditions that are difficult to simulate in the laboratory. Many of these naturally occurring materials are finding use in fields like optics, electronics, mechanics, or materials science. A most intriguing example is diamond, which is formed in the lithosphere at considerable depth, supplying the required high pressure and temperature.[2] Furthermore, quasi-crystals, whose discovery was awarded by the Nobel Prize in 2011,[3] were very recently found in sky-fallen meteorites,[4] although they were hitherto believed to be only synthetically accessible. A further notable natural compound is graphite as a source for graphene, the thinnest atomic sheet isolated to date.[5] Owing to its linear energy dispersion relation at low carrier energies, graphene has emerged as one of the most promising component of future electronic devices. Closely related to graphene is a new class of materials called topological insulators (TIs) whose surface states not only display a linear dispersion but also spin polarization due to spin-momentum locking.[6] The helical surface states of TIs could provide access to novel, fascinating physical phenomena such as magnetic monopoles[7] or Majorana fermions.[8] Moreover, the spin polarized surface electrons are of interest for spintronic applications.[9] Very recently, synthetic TIs of optimized composition have been demonstrated to exhibit an ambipolar gating effect as an indicator of predominant charge transport through the surface states.[10]

Quartz veins, which have a long history in gold mining, represent a promising location to encounter TIs in nature. In fact, chalcogenides have been shaped in such veins over billions of years via hydrothermal deposition at high temperatures (>300°C).[11] Some of these compounds contain heavy elements like bismuth or antimony which are a necessary ingredient for the strong spin-orbit coupling (SOC) operative in TIs. One prospective candidate is Kawazulite, a metallic gray mineral with the general chemical composition $Bi_2(Te,Se)_2(Se,S)$, which is named after the Kawazu mine in Japan, where it was first discovered. It belongs to the Tetradymite group (rhombohedral, space group $3\overline{R}m$) featuring a crystal structure composed of quintuple layers (VI(1) – V – VI(2) – V – VI (1), where VI = (Se, Te, S) and V = Bi) that are held together through van-der-Waals bonds. Its crystal structure was first reported by Bland *et al.* in 1961.[12]

The samples investigated in this work originate from a former gold mine in Jilove, Czech Republic. **Figure 1a** shows an optical micrograph of one of the investigated specimens (purchased from Mineralienkontor, Germany), comprising Kawazulite crystals as metallic-like inclusions in a quartz matrix. Chemical analysis of mechanically extracted, tiny Kawazulite pieces by inductively coupled plasma mass spectrometry (ICP) revealed a composition of $(Bi_{2.12}Sb_{0.06})Te_2(Se_{0.14}S_{0.32})$. Besides these major elements, fourteen additional trace elements could be identified (for a detailed list and supporting XPS and EDX data, see Supplementary Information). Powder diffractometry (see **Figure 1b**) indicated that the Kawazulite is composed of two different crystalline phases, which both crystallize in rhombohedral ($3\overline{R}m$) structure and exhibit slightly different lattice constants (see Methods for details). While the major phase (89.6 wt%) is characterized by the lattice parameters $a$ = 4.25 Å and $c$ = 29.70 Å (Se rich), with a stoichiometry of $Bi_2(Se_{0.2}S_{0.74})(Te_{0.61}Se_{0.39})_2$, values of $a$ = 4.37 Å and $c$ = 30.42 Å (Se rich) and a stoichiometry of $Bi_2(Se_{0.59}Te_{0.41})Te_2$ were found for the minor phase (~10.4 wt%). These lattice parameters are smaller than for $Bi_2Te_3$, consistent with the fact that Sb and Se atoms are smaller than Bi and Te. Moreover, the observation of (107) and (00.12) peaks (see **Figure 1c**) testifies a sizeable extent of chalcogen ordering and correspondingly the presence of an ordered layer structure[13] (see sketch of the unit cell in **Figure 1d**). In addition to the two Kawazulite phases, some spectra also contain peaks originating from the surrounding matrix of the mineral, which could be identified as Clinochlore $(Mg,Fe,Al)_6(Si,Al)_4O_{10}(OH)_8$ (in good agreement with the determined elemental composition, see Supplementary Information).

The layered structure of the Kawazulite is confirmed by its Raman spectrum, which displays two pronounced peaks located at 105 cm$^{-1}$ and 151 cm$^{-1}$ (**Figure 1e**). Similar to the layered chalcogenides $Bi_2Se_3$, $Bi_2Te_3$, $Sb_2Te_3$ and their alloys, these two peaks can be assigned to the $E^2_g$ and the $A^2_{1g}$ peak, respectively, which originate from the opposite phase motion of the outer V and VI(1) atoms of the quintuple layers. Based upon the dependence of the $E^2_g$ peak position on the chemical composition of the alloys,[14] it can be concluded that the Kawazulite comprises a Te-rich $Bi_2(Te_{1-x}Se_x)_3$ matrix, with a Se content $x < 0.6$. Further clues regarding the composition derive from the absence of a split of the $A^2_{1g}$ peak. Replacing the heavy Te atoms by Se atoms is documented to shift this peak from 134 cm$^{-1}$ in pure $Bi_2Te_3$ to higher energy, with a split into two components at 139 cm$^{-1}$ and 150 cm$^{-1}$ upon reaching $Bi_2Te_2Se$ stoichiometry.[14] It hence follows that the Se concentration in Kawazulite is slightly below $x = 1/3$. The quite high energy of the $A^2_{1g}$ peak (151 cm$^{-1}$) can be explained by the Sb content of the sample, in close correspondence to the up-shift detected upon replacing the heavy Bi atoms in $Bi_2Te_3$ by lighter Sb atoms.[14]

The electronic structure of the (111)-surface near the Γ-point, determined by angle-resolved photoelectron spectroscopy (ARPES) of a tiny (0.7 x 0.7 mm²) Kawazulite crystal, is displayed in **Figure 2a**. A surface state with the typical Dirac-like conical dispersion can be seen in the center, which verifies the natural Kawazulite to be a TI. In **Figure 2b**, the corresponding electronic band structure is schematically illustrated. The general dispersion closely resembles that of the well investigated Bi chalcogenide family. While the area of enhanced intensity above a binding energy of 0.4 eV can be attributed to the bulk valence band, the bulk conduction band is difficult to discern near the Fermi level presumably due to a suppression of the spectral intensity through matrix element effects or sample/surface quality. If the Fermi level actually resides within the occupied states, comparison with similar compounds suggests that it is less than 0.1 eV away from the band edge. In any case, it can be concluded that the size of the band gap is larger than 0.25 to 0.35 eV, the precise magnitude depending on whether the conduction band is slightly occupied or not. The latter value exceeds the band gap reported for pure $Bi_2Te_3$,[16] which may be explained by the higher electronegativity of sulfur vs. tellurium.[17]

The appreciable sulfur content of the Kawazulite is furthermore notable in view of the finding that replacement of Se by S atoms in $TlBi(S_{1-x}Se_x)_2$ weakens the spin-orbit interaction in these compounds, which causes a topological quantum phase transition from a topological state (Se rich) to a trivial insulating state (S rich).[15] In the topological insulators $Bi_2X_3$ (X = Se,Te), a similar transition takes place, which is, however, coupled to a structural phase transition. For increasing sulfur content in $Bi_2Se_3$ a change from rhombohedral $R\bar{3}m$ (Se rich) to orthorhombic crystal structure *Pnma* (S rich) occurs, resulting in a trivial band insulator with a large gap.[16] In $Bi_2(Te_{1-y}S_y)_3$, by contrast, the heavier Te stabilizes the rhombohedral structure up to high sulfur contents of $y = 0.5$.[16] As natural Kawazulite contains a sizeable amount of Te, a similar stabilization can be expected, consistent with the rhombohedral crystal structure revealed by XRD analysis.

Evaluation of the ARPES data yields a Fermi vector of $k_F = 0.091$ Å$^{-1}$ ± 0.002 Å$^{-1}$, a Fermi velocity of $v_F = 8.63 \cdot 10^5$ m/s, and (by assuming a circular Fermi surface) a surface electron density of $n_s = k_F^2 / 4\pi = 6.45 \cdot 10^{16}$ m$^{-2}$. Furthermore, we determined the electrical conductivity of individual thin flakes prepared by micromechanical cleavage of the extracted crystals (see inset of **Figure 3a** for an optical image of a typical device in Hall bar configuration, or **Figure 3b** for an AFM image of a ribbon-like device). Thus obtained micro-flakes exhibit the bulk crystal structure, as evidenced by the transmission electron (TEM) and selected area diffraction (SAD) data in **Figure 3c and d**, respectively. From the SAD pattern, it furthermore follows that they reproducibly comprise (111) crystal faces which are oriented parallel to the substrate surface. In **Figure 3a**, the Hall resistance of a 12 nm thick flake is plotted as a function of the applied *B*-field. From low temperature Hall measurements on a range of different flakes, a Hall mobility between 300 and 1300 cm²/Vs, and an average electron density of $n = 1 \cdot 10^{17}$ m$^{-2}$ (normalized by the sample thickness) could be derived (see Supplementary Information for details). The latter density is only slightly higher than the value gained from the ARPES data, indicative of a dominant contribution of the surface state to the total charge transport.

To further consolidate this assertion, we evaluated the low temperature magnetoresistance of the samples at different tilting angles θ. As apparent from **Figure 4a**, where data gained from the device in **Figure 3b** are shown, there emerges a pronounced weak antilocalization (WAL) peak in the sheet conductance $\Delta\sigma = \sigma(B) - \sigma(B=0)$. Such behavior can be understood based upon the Berry phase difference of π between the electron wavefunctions belonging to time reversed paths in a diffusive conductor with strong spin orbit coupling.[20] The resulting constructive interference along these paths prevents carrier localization, leading to a positive correction to the total conductivity. However, when the time reversal symmetry is broken by applying a magnetic field, antilocalization is no longer effective and correspondingly the conductivity decreases. When $\Delta\sigma$ is plotted in dependence of $B\cdot\cos(\theta)$, *i.e.*, the *B*-field component normal to the surface, all curves coincide (see **Figure 4b**), signifying the 2D character of the effect.[21] In addition, the low-field magnetoconductance data can be well-fitted by the Hikami-Larkin-Nagaoka (HLN) model[22] for 2D localization, according to:

$$\Delta\sigma(B) = \alpha(e^2/h)[\ln(B_\phi/B) - \Psi(1/2 + B_\phi/B)], \tag{1}$$

where $\Psi$ is the digamma function, $l_\phi$ the phase coherence length, $B_\phi = \hbar/(4el_\phi^2)$ and the constant $\alpha$ provides information about the nature of the transport channel (for a bulk sample with a single conductive surface state $\alpha = -1/2$). Previous studies on various types of TIs have found $\alpha$ values between -0.3 and -1.1, which were interpreted as evidence for transport through a single surface state, two decoupled surface channels with comparable phase coherence length,[24] or intermixing of bulk and surface conductance.[25] By fitting the low-field magnetoconductance of the above sample at different temperatures (see **Figure S2b**), a phase coherence length between 200 nm at 1.5 K and 60 nm at 30 K was obtained. In addition, the fits yielded an average value of $\alpha = -0.92$, which suggests electrical transport through two decoupled surface channels, *i.e.*, the top and bottom surface with comparable phase coherence length. As apparent from **Figure 4c**, the phase coherence length displays a pronounced decrease with increasing temperature, which is attributable to enhanced electron-phonon and electron-electron interactions.[26] Assuming that electron-electron scattering is the dominant mechanism, we fit the data by a simple power law, where $l_\phi$ is proportional to $T^{-1/2}$ for a 2D system, and $\propto T^{-2/3}$ in the 3D case.[27] From such fit, we obtain an exponent very close to -1/2, thus further corroborating the 2D nature of the WAL in the present samples.

Another intriguing observation is the emergence of pronounced universal conductance fluctuations (UCF) in the magnetoconductance curves,[28] as exemplified in **Figure 4a**. These oscillations are non-periodic, but perfectly reproducible and symmetric in *B*-field, and provide valuable information about the defect distribution. It is evident from the magnetoconductance plot in **Figure 4a** that the UCF amplitude decreases substantially with increasing temperature. According to theory,[28] the root mean square (rms) of the UCF is related to the phase coherence length of the electrons via $\text{rms}(\delta G) \sim (l_\phi / L)^{(4-d)/2}$, where *L* is the typical edge length of the sample and *d* the dimensionality of the electron system. On this basis, the dimensionality of the observed UCF can be determined by plotting $\text{rms}(\delta G)$ as a function of $l_\phi$ derived from the WAL measurements (see **Figure 4d**). The very good quality of the fit by a straight line ($d = 2$) signifies a 2D character of the UCF. To further consolidate this conclusion, we studied the angle dependence of the magnetoconductance (for details see Supplementary information). In **Figure 4e**, the position of two selected peaks (see guideline to the eye in **Figure S3b**) is plotted versus the tilting angle θ. That the data can be well fitted by a $1/\cos(\theta)$ function indicates that the peak positions depend only on the *B*-field component normal to the substrate plane, thus confirming the 2D character of the charge transport.

The discovery of Kawazulite as a natural TI whose electrical properties are comparable to those of state-of-the art synthetic compounds renders it likely that further minerals belonging to this fascinating class of materials can be located in nature. Prospective candidates are for instance the members of the Tetradymite and Aleksite group which together comprise more than twenty compounds. One useful search guideline may be established on the basis of theoretical predictions, making use of, e.g., high-throughput robustness descriptors.[29] This could involve sorting out those compounds which occur as minerals, and further select those exhibiting high defect formation energies. Due to their geological age, the crystal structure of these minerals should have reached thermodynamic equilibrium and therefore an ultimately low defect concentration. In this manner, it may be possible to spot natural TIs which display further reduced bulk doping and accordingly even better accessible surface state transport, as compared to Kawazulite. In addition, the very recent visualization of topological phases in photonic quasi-crystals[30] renders it likely that also their natural counterparts can display "topological" properties.

**Experimental section**

X-ray powder diffraction data of the powdered Kawazulite sample were collected at room temperature with a Stoe Stadi-P transmission diffractometer (primary beam Johann-type Ge monochromator for Ag-K$\alpha_1$-radiation, linear PSD) with the sample sealed in a glass capillary of 0.5 mm diameter (Hilgenberg, glass No. 50). This geometry was chosen since preliminary powder patterns recorded on a Bragg-Brentano diffractometer exhibited strong preferred orientation even after careful sample preparation. The powder pattern was recorded for 20 h in the range from $2\theta = 2\text{-}45°$ with a step width of $2\theta = 0.01°$ using a linear position sensitive detector with an opening of approximately $2\theta = 12°$. The sample was spun during measurement for better particle statistics. Further details on the experiments and data evaluation are provided in the Supplementary Information.

The ARPES measurements were carried out with a hemispherical SPECS HSA3500 electron analyzer with an energy resolution of ~10 meV. Monochromatic HeI (21.2 eV) was used as a photon source in the experiments performed at a pressure below $3\cdot10^{-10}$ mbar. The samples consisted of small, monocrystalline grains (approximately 0.7 x 0.7 mm$^2$ in size) extracted from the mineral specimens. Prior to the ARPES measurement, the tiny crystals were cleaved in the load lock at a pressure of $2\cdot10^{-7}$ mbar.

Toward electrical characterization, Kawazulite was subjected to micromechanical cleavage using a Scotch tape, and flakes with a thickness up to a few tens of nanometers were selected. Individual flakes were provided with Ti(4 nm)/Au(100 nm) contacts in Hall-bar or van-der-Pauw geometry by standard e-beam lithography. The contact regions were treated with 50 s Ar plasma before evaporation of the metal contacts in order to reduce the contact resistance.[31] Electrical measurements were performed in an Oxford cryostat at the base temperature of 1.3 K and magnetic fields up to 12 T.

Transmission electron microscopy (TEM) and selected area diffraction (SAD) measurements were carried out with a Philips CM 200 operated at 200 kV. The samples for TEM investigations were prepared by micromechanical exfoliation of Kawazulite onto a TEM grid.

FIGURES

**Figure 1.** X-ray and Raman characterization of Kawazulite. (a) Optical micrograph of the investigated specimens with a size between 3 and 10 mm. (b) Rietveld plots for the two Kawazulite phases. (c) Magnified plot of the X-ray diffraction (XRD) data within the range of 11° to 13.4°. (d) Sketch of the unit cell of the mineral, as derived from the XRD analysis. (e) Raman spectrum recorded from a Kawazulite crystallite ($\lambda_{exc}$ = 633 nm). The two blue lines are Lorentzian fits, while the red line corresponds to the total fit.

**Figure 2.** ARPES of Kawazulite. Raw data (a) and sketch of the electronic band structure derived from the measurement (b).

**Figure 3.** Hall measurements on Kawazulite. (a) Hall data gained from the device shown in the inset (optical micrograph). (b) Atomic force microscopy image of the device discussed in the main text (scale bar corresponds to 1 µm). The height profile is taken along the dashed line. (c) TEM image and (d) SAD pattern of a typical micro flake.

**Figure 4.** Weak antilocalization and universal conductance fluctuations. (a) Magnetoconductance recorded at different temperatures. (b) Low $B$-field range of the data shown in panel (a), after normalization to the $B$-field component normal to the surface. (c) Temperature dependence of the phase coherence length extracted from the fits in figure S2 (b) (see Supplementary Information). The curve fit (red line) indicates scaling of the phase coherence length with $T^{-0.49}$. (d) Root mean square of the UCF as a function of the phase coherence length calculated from the WAL fits. The red line is a linear fit to the data. (e) B-field position of two characteristic features of the UCF in dependence of the tilting angle of the sample. The solid lines are $(1/\cos\theta)$ fits to the data.

ASSOCIATED CONTENT

**Supporting Information**.

Results of the semi-quantitative ICP analysis; Details on the XRD results; XPS data of the sample; Additional electrical transport data; Additional WAL data; Details on UCF evaluation. This material is available free of charge via the Internet at http://pubs.acs.org.

AUTHOR INFORMATION

**Corresponding Author**

*P.G.: E-mail: p.gehring@fkf.mpg.de

**Author Contributions**

P.G. developed the basic idea. The experimental activities were guided by P.G., M.B. and K.K. H.M.B. and C.R.A. performed and interpreted the ARPES measurements. R.D. did XRD measurements. Y.W. performed TEM measurements. P.G. performed electrical transport measurements. All authors contributed to the content of the manuscript.

ACKNOWLEDGMENT

The authors are grateful to A. Leineweber for valuable discussions on the X-ray data, as well as M.-L. Schreiber for help with the chemical analysis. M. Konuma is acknowledged for support with the XPS measurements. C.R.A. acknowledges support by the Emmy Noether program. We are grateful to S. Moeckel for providing valuable information about origin and composition of the mineral samples, as well as to P. van Aken for supporting the TEM experiments.

**Figure 1.**

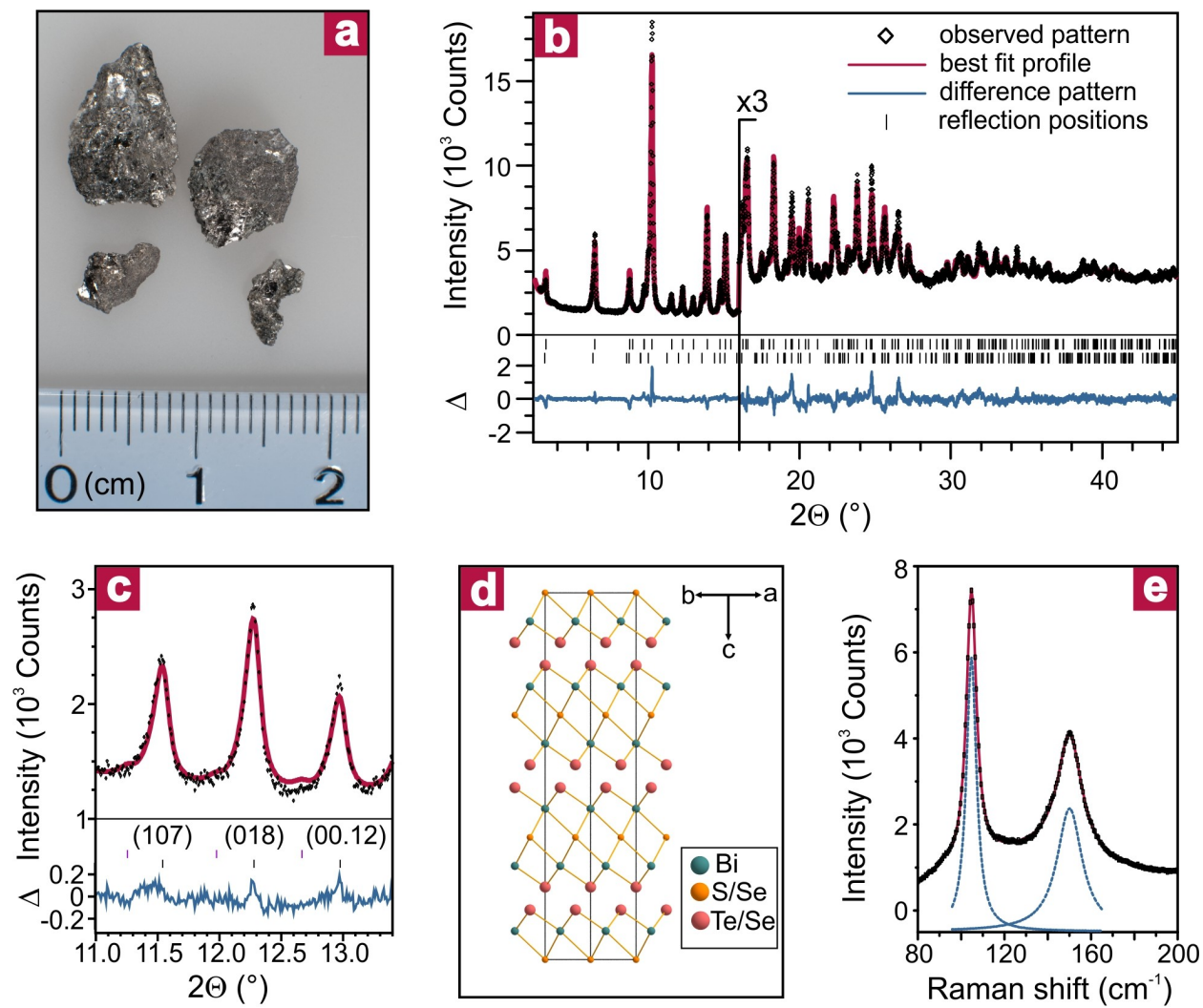

**Figure 2.**

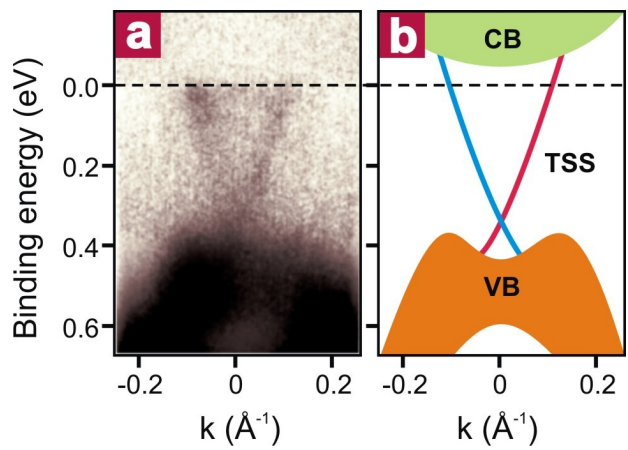

**Figure 3.**

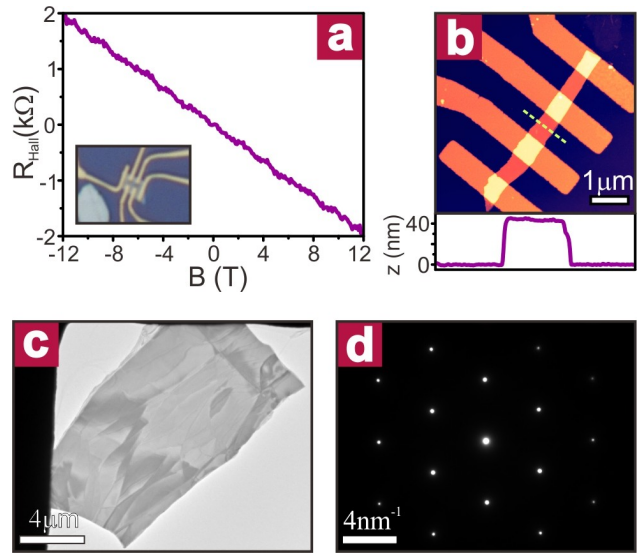

**Figure 4.**

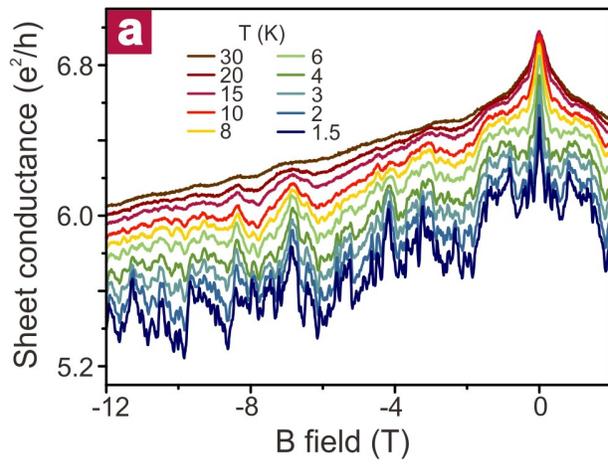
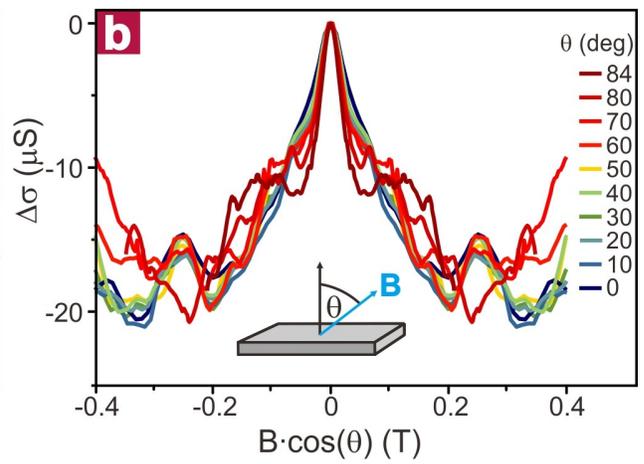
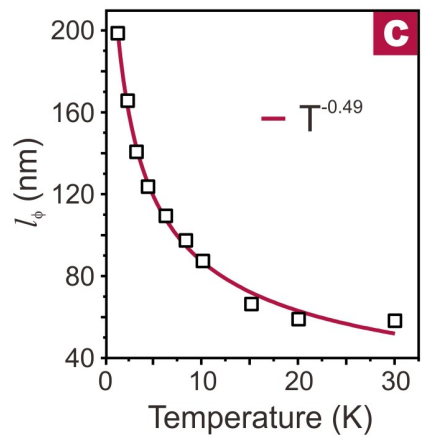
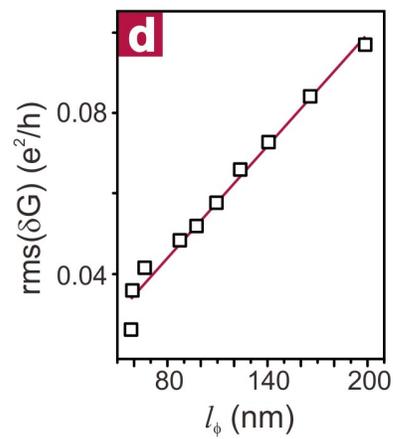
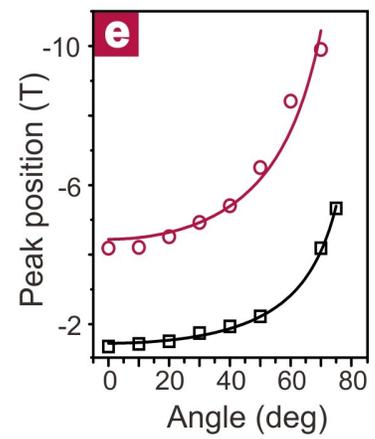